# An Experimental Investigation of Secure Communication With Chaos Masking

Sourav Dhar[1] and Kabir Chakraborty[2]

[1]Department of Information Technology

Pailan College of management And Technology; P.O- Pailan ; Kolkata-700104

dhar.sourav@gmail.com

[2]Department of Electrical Engineering

Tripura University; Suryamani Nagar; Tripura (West); Pin-799130

kabir_jishu@rediffmail.com

**Key Words:** Buck Converter, Boost Converter, Chaotic Signal.

**Abstract:**

The most exciting recent development in nonlinear dynamics is realization that chaos can be useful. One application involves "Secure Communication". Two piecewise linear systems with switching nonlinearities have been taken as chaos generators. In the present work the phenomenon of secure communication with chaos masking has been investigated experimentally. In this investigation chaos which is generated from two chaos generators is masked with the massage signal to be transmitted, thus makes communication is more secure.

## 1.     Introduction:

Chaos is a periodic long-term behavior in a deterministic system. It exhibits a sensitive dependence of a system's dynamical variable on the initial conditions meaning that no two chaotic systems will evolve in the same way. Trajectories of two perfectly identical chaotic systems starting with nearby initial conditions diverge from each other exponentially. Synchronization means that the trajectories of two chaotic systems be locked to each other. Hence synchronization seems unlikely for two chaotic systems if trajectories start from initial conditions that differ slightly. Moreover, in practical applications the existence of noise (both internal and external) and system imperfect identification makes the hope of synchronizing two chaotic systems even more complicated. Nonetheless, it has been established that [1], [2], [3] synchronization of chaotic dynamical systems is not only possible but it is believed to have potential applications in communication. The strategy is that when we transmit the message to a friend, we mask it with louder chaos. An outside listener only hears the chaos, which sounds like meaningless noise. But if the friend has a magic receiver that perfectly reproduces the chaos, then he can subtract off the chaotic mask and listen to the message. This synchronization is possible only when a similar chaotic circuit as that of sending end is fabricated. If the configuration circuit is secret, it is impossible to extract information from the transmitted message.

Hence there has been growing interest in the possibility of synchronizing chaotic signals. This idea has been tested theoretically as well as experimentally in the variety of linear dynamical system including Chua's circuit and Driven Chua circuit [4]. Lai et al [5] demonstrated that applying small temporal parameter perturbation to one of them could synchronize two identical chaotic systems. But in all these methods, synchronization is possible without any massage signal. Murli and Laksmanan [6] investigated the method of transmitting signal using chaos synchronization in Vander Pol-Duffing oscillator. But they transmitted very weak signal with chaotic masking. The synchronization is failed with high strength of massage signal. These shortcomings have been overcome by Chakraborty et al [7], in their work, the synchronization and faithful recovery of massage at the receiving end is independent of massage signal strength. In that work ,buck converter [7] has been taken as chaos generator.

In the present work, to make communication more secured, buck converter and boost converter are chosen as chaos generator, both chaotic signal are synchronized and masked with massage signal.

## 2. Buck Converter Circuit:

Buck Converter circuit is generally a chopper circuit that modulates input DC to output DC voltage at different levels. The input voltage is connected in series with an inductor and load resistance. There are two switches (i) S is a controlled switch (realized by a transistor) and (ii) D is uncontrolled (realized by a diode). The controlled switch is in series with the input voltage and the uncontrolled switch is in parallel with the combination of inductor and load resistor. A capacitor is connected to smoothen the load voltage waveform (shown by the dash line in Fig.1). This is optional, and is not considered in the flowing analysis for the sake of simplicity.

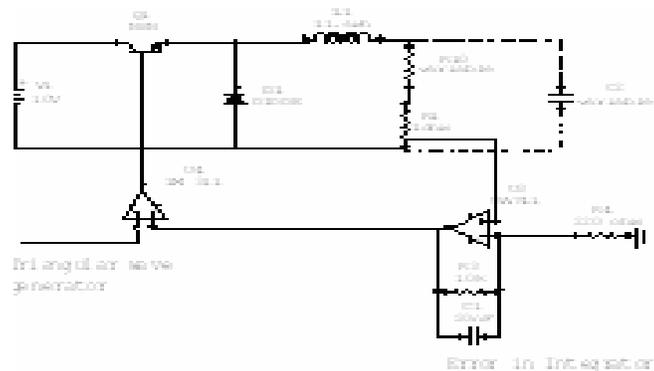

Fig.1. Buck Converter circuit.

When S is in the ON state, the diode is OFF and the current builds up through the inductor and load. When the transistor is OFF the current freewheels through the diode and decays. By properly choosing the switching frequency and inductor value, one achieves continuous current operation and the average output voltage is then given by,

$$V_{out} = V_{in} \frac{T_{on}}{T_{on} + T_{off}} \quad \text{---------------(1)}$$

Where $V_{in}$ = input voltage, $V_{out}$ = output voltage, $T_{on}$ and $T_{off}$ = on and off times of the controlled switch.

## 2.1 Switching control:

To control the switching through current feedback, the load current is sensed as the voltage across 1 ohm resistance connected in series with the load. The signal is integrated by passing it through a low pass filter. The output of the integrator is compared with a triangular wave voltage. The output of the comparator that compares the triangular wave and error integrator voltage is used to control the on and off period of the switch. When error integrator voltage is less than triangular wave voltage, the controlled switch is turned on, and current flows through the inductor and load. The controlled switch is turned off when error integrator voltage is greater than triangular wave voltage. At this instant the voltage across the inductor reverses its polarity and the current path completes through the diode and load. Consequently the stored energy in the inductor decreases and output voltage falls. When the integrator output voltage drops below the triangular wave voltage, the switch is turned on and the process continues. The choke is designed for linear inductance and a diode with negligible storage is employed.

## 2.2 System Analysis:

The system is governed by two sets of linear differential equation pertaining to the ON and OFF states of the controlled switch. The error integrator voltage and load current are taken as state variables.

During the on period, the equations are:

$$\frac{di}{dt} = \frac{V_{in}}{L} - \frac{R_l}{L} i \quad \text{-----------------------------(2)}$$

$$\frac{dv_{io}}{dt} = i\left(\frac{R_i + R_2}{CR_1 R_2} - \frac{R_l}{L}\right) - \frac{v_{io}}{R_1 C} + \frac{V_{in}}{L} \quad \text{----------(3)}$$

During the off period, the state equations are

$$\frac{di}{dt} = -\frac{R_l}{L} i \quad \text{-----------------------------(4)}$$

$$\frac{dv_{io}}{dt} = i\left(\frac{R_i + R_2}{CR_1 R_2} - \frac{R_l}{L}\right) - \frac{v_{io}}{R_1 C} \quad \text{--------------(5)}$$

Though the equations are linear, bifurcation [7] and chaos appear in this circuit because of the switching nonlinearity.

## 3. Boost converter circuit:

Boost converter circuit like the buck converter circuit [Fig.2], is a chopper circuit that steps up the output DC voltage ($V_o$) level as compared to its input voltage ($V_{in}$) level, that is,

$$Vo = k / (1-k) V_{in} \quad \text{--------------(6)}$$

where, 'k' is the duty ratio of the switch. In this project work, a current controlled boost converter operating in continuous current mode is considered.

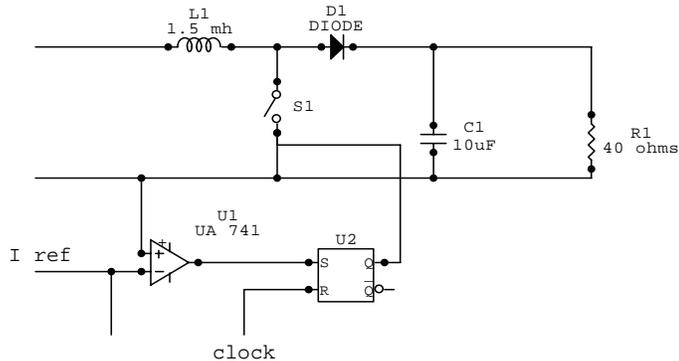

Fig. 2 Boost converter circuit with current control feed back

### 3.1. Switching control:

This control logic the switch turned on by the clock pulses that are spaced T seconds apart. When the switch is closed, the current increases until it reach the specified reference value. The switch opens when I= $I_{ref}$. Any clock pulse arriving during ON period is ignored. Once the switch has opened, the next clock pulse causes it to close.

### 3.2. System Analysis:

The evolution of the state variable I and $v_c$ during the ON and OFF periods are described by the differential equations below

During ON period,

$$\frac{di}{dt} = \frac{V_{in}}{L} \quad \text{-----------(7)}$$

$$\frac{dv_c}{dt} = \frac{v_c}{CR} \quad \text{----------(8)}$$

During OFF period,

$$\frac{di}{dt} = \frac{V_{in}}{L} - \frac{v_c}{L} \quad \text{----------(9)}$$

$$\frac{dv_c}{dt} = \frac{I}{C} - \frac{v_c}{CR} \quad \text{---------(10)}$$

These are linear equations but chaos is due to switching nonlinearity.

## 4. Communication with masking:

In Fig.3, the massage signal s (t) is to be transmitted is added with output of error integrator and the output of the boost converter and buck converter. The resulting signals X (1) + X (2) + s (t) is used to compare the triangular wave at the transmitter end (Fig3). The same signal is transmitted through the communication channel. Now, the message signal s (t) is masked with chaotic signals

X (1) and X (2). If one tries to get the message, he can not extract it because he does not know for what values of the parameters chaos appears in the buck and boost converters.

At the receiving end (Fig.3), the transmitted signals X (1) +X (2) + s (t) is fed to the inverting input of the comparator. The circuit parameters and configuration of the buck and boost converter is same as that of the transmitter. Due to synchronization the receiver buck converter and boost converter also generates X (1) and X (2) respectively. Now subtracting X (1) and X (2) form X (1) + X (2) + s (t), the message signal s (t) is obtained.

## 5. Results:

The proposed scheme is verified experimentally for sinusoidal wave and triangular wave. The results are shown in fig.4.1, 4.2, 4.3, and 4.4.

## 6. Conclusion:

Earlier it was developed that message can be masked with chaos generated from one circuit i.e. from buck converter. In my circuit the chaotic signal is generated from two circuits i.e. actually more chaotic. So it is impossible to determine the chaotic signal for any person. Thus the communication is said to be "secure". The work can be proceed further by considering the effect of white Gaussian noise.

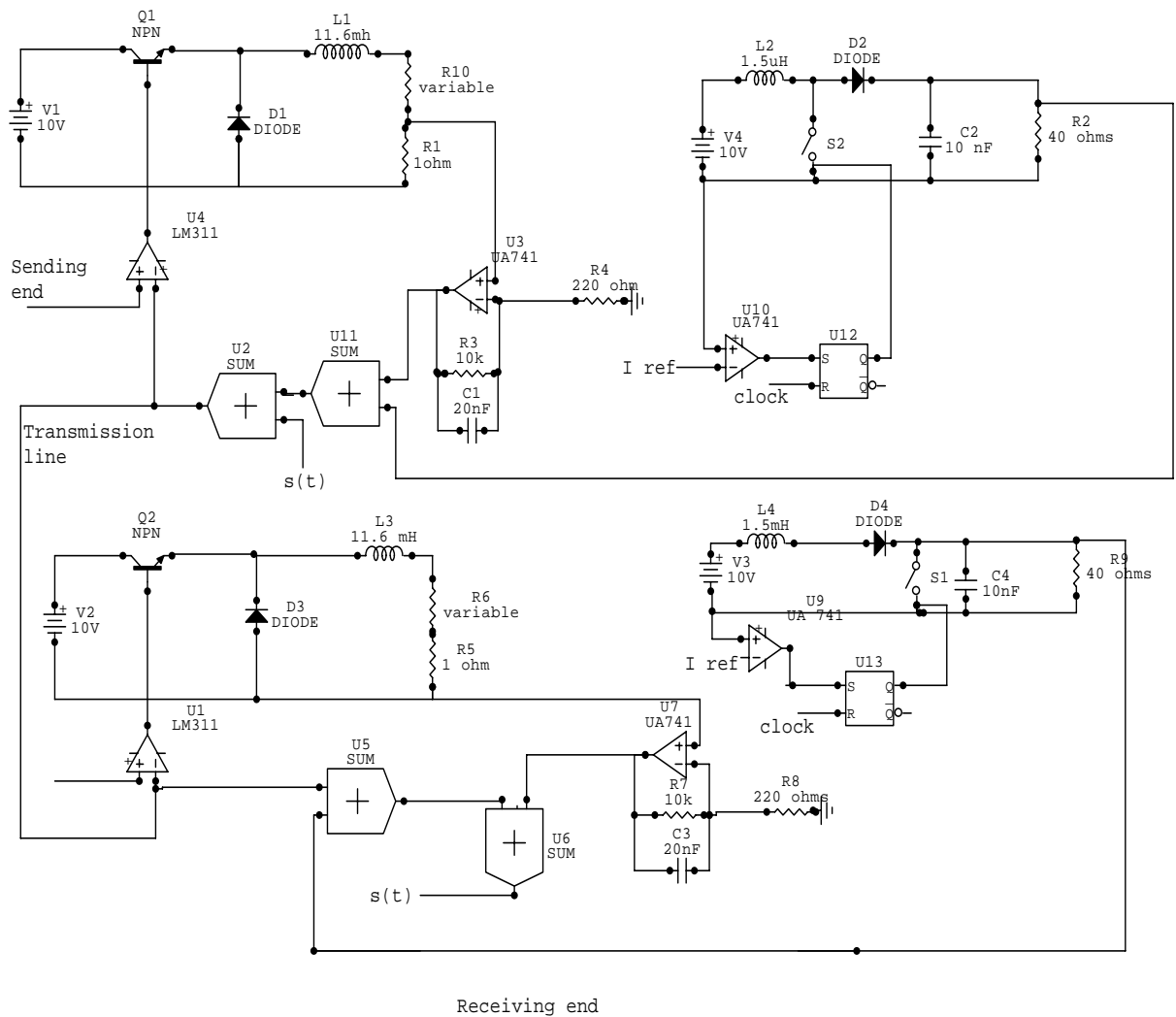

Fig.3 Secure communication Scheme

## 7.    Acknowledgement:

The authors are thankful to Prof. A.K. Mukhapadhyay, Vice Chancellor, Tripura University and Prof. Shouvic Roy, HOD, IT Department ,PCMT for providing their support both technically and financially.

**Program, result**

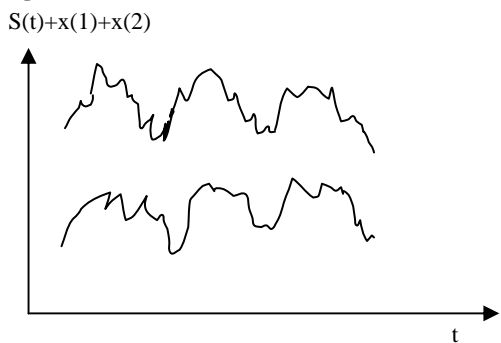

Fig. 4.1 Transmitted masked sine wave.

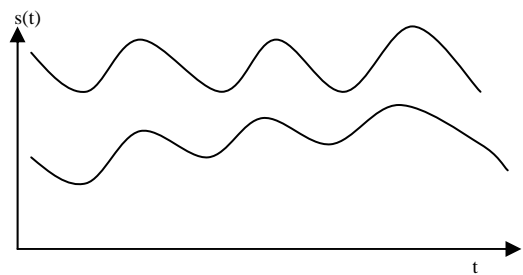

Fig 4.2 Received sine wave

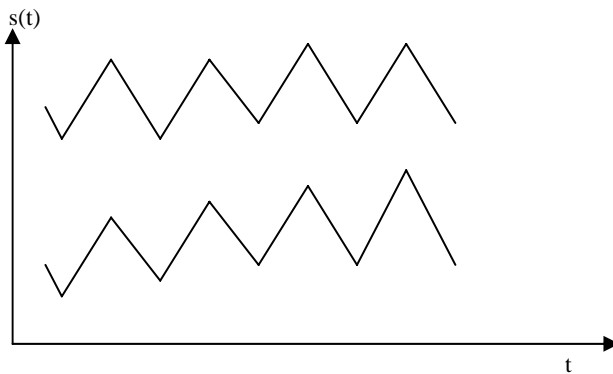

Fig 4.4 Received triangular wave

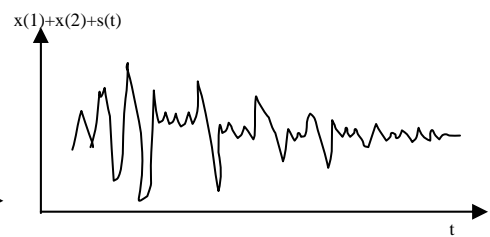

Fig4.3 Transmitted masked triangular wave